\documentclass[aps,prl,twocolumn,showpacs,showkeys,pacsgroupedaddress]{revtex4-1}
\usepackage[english]{babel}
\usepackage[dviwin]{graphicx}
\usepackage[dvips]{color}
\usepackage{graphicx}% Include figure files
\usepackage{dcolumn}% Align table columns on decimal point
\usepackage{bm}% bold math
\usepackage{amsmath}
\begin{document}
% Use the \preprint command to place your local institutional report
% number in the upper righthand corner of the title page in preprint mode.
% Multiple \preprint commands are allowed.
% Use the 'preprintnumbers' class option to override journal defaults
% to display numbers if necessary
%\preprint{}
%
%Title of paper
%
\title{  Quantum speed limit time in a magnetic resonance}
%\title{Visualization of Rabi oscillations  in a magnetic &resonance: quantum speed limit in  qudits}
%
% repeat the \author .. \affiliation  etc. as needed
% \email, \thanks, \homepage, \altaffiliation all apply to the current
% author. Explanatory text should go in the []'s, actual e-mail
% address or url should go in the {}'s for \email and \homepage.
% Please use the appropriate macro foreach each type of information

% \affiliation command applies to all authors since the last
% \affiliation command. The \affiliation command should follow the
% other information
% \affiliation can be followed by \email, \homepage, \thanks as well.
\author{E. A. Ivanchenko}
\email{eaivanchenko1@gmail.com,\,yevgeny@kipt.kharkov.ua}
%\homepage[]{Your web page}
%\thanks{}
\affiliation{Institute for Theoretical Physics, National Science
Center \textquotedblleft{}Institute of Physics and
Technology\textquotedblright{},
 \\
  1, Akademicheskaya
str., 61108 Kharkov, Ukraine}
%Collaboration name if desired (requires use of superscriptaddress
%option in \documentclass). \noaffiliation is required (may also be
%used with the \author command).
%\collaboration can be followed by \email, \homepage, \thanks as well.
%\collaboration{}
%\noaffiliation
\date{\today}
\begin{abstract}
 A visualization  for dynamics of a qudit spin vector in a time-dependent magnetic field is realized by means of mapping a solution for a spin vector on the three-dimensional spherical curve (vector hodograph).
The  obtained results  obviously display the  quantum interference of precessional and nutational effects on the spin vector in the magnetic resonance.  For any  spin the bottom bounds of the   quantum speed limit time (QSL) are found. It is shown  that the    bottom bound  goes down when using  multilevel spin systems.
Under certain conditions the non-nil minimal time, which is  necessary to achieve the orthogonal state  from  the initial one,  is attained at  spin S=2.
An estimation of the product of two and three standard deviations of the
spin components are presented.
 We discuss the dynamics of  the mutual uncertainty, conditional uncertainty  and  conditional variance in terms of spin standard deviations.
The study can find  practical applications in the magnetic resonance, 3D visualization of computational data and in  designing  of optimized information processing
devices for quantum computation and communication.
\end{abstract}
% insert suggested PACS numbers in braces on next line
\pacs{87.63.L,82.56.-b,03.67.-a}
%Quantum systems with finite Hilbert space, 03.65.Aa
\keywords{3D visualization, spin  resonance,
%\,Quantum information,visibility,
quantum speed limit time, Mandelstam-Tamm bound}
% \pacs {Visual imaging 87.63.L-, Visual perception,
% 42.66.Si, Nuclear magnetic resonance (NMR)
%in chemical physics, 82.56.-b,
% 03.67.-a Quantum information}
 % Visual imaging, 87.63.L-Visual imaging
%\keywords: Visualization, NMR, Quantum information
% insert suggested keywords - APS authors don't need to do this
%\maketitle must follow title, authors, abstract, \pacs, and \keywords
\maketitle
% body of paper here - Use proper section commands
% References should be done using the \cite, \ref, and \label commands
%
%\section{Introduction}\label{I}
%\textit {Introduction}\label{I}
%
%\textit {Introduction.}\label{I} Imaging and visualization are among the most dynamic and
%innovative  research areas of  the past  decades.
% This activity arises from the requirements of
%important practical applications such as the computational data
%visualization, the medical images processing   for assisting
%medical diagnosis and intervention, and the 3D geometry
%reconstruction and processing for computer simulations.
%Due to the development of more powerful hardware
%resources, mathematical and physical methods, investigators have
%been incorporating advanced computational techniques to derive
% methodologies which can better   solve
%the problems encountered. \\
%\indent In this paper we are going to introduce a 3D visualization %scheme   of the spin qudit  vector evolution  in a magnetic field.
%a paramagnetic resonance in a continuous mode
\section{Introduction} %As it is known,
The magnetic resonance realization in a continuous mode depends on the kind of  magnetic field modulations. Let's consider the spin dynamics in the alternating
field \cite{PhysicaBIvanchenko,StabilizationIvanchenko}
\begin{equation}\label{eq:1}
\vec{h}(t)=\left( h_1 \mathrm{cn}(\omega t|k),\;h_2 \mathrm{sn}
(\omega t|k),\;H \mathrm{dn}(\omega t|k)\right),
\end{equation}
where $\mathrm{cn},\mathrm{sn},\mathrm{dn}$ are the Jacobi
elliptic functions \cite{AbramovitzStegun}, $\omega$ is the field frequency. Such field modulation
under the changing of the elliptic modulus $k$ from 0 to 1
describes the whole class of field forms from
trigonometric \cite{IIRabi} ($k=0$)
to the exponentially impulse ones ($k=1$)
\cite{BambiniBerman}. The
elliptic functions $\mathrm{cn}(\omega t|k)$ and$\;
\mathrm{sn}(\omega t|k)$ have a real period $4K/\omega$, while the function $\mathrm{dn}(\omega t|k)$ has a period of
half a duration. Here $K$ is a full elliptic integral of the
first kind \cite{AbramovitzStegun}.
At $k\neq0$ and $h_1=h_2=h $ we call such field consistent (cf).
 At $k=0$ and $h_1=h_2=h $ it is a circularly polarized magnetic field
 \cite {IIRabi}, where the scalar  of the vector of the magnetic field does not depend on  time (a rotating magnetic field), and at $k=0$ and
$h_1=h, h_2=0$ it is a linearly polarized field \cite{BlochSiegert}.\\
%\indent In the paper
%\cite{FeynmanVernonHellwarth} real  functions have been constructed  from  the Schrodinger equation solutions, which have direct physical sense and
%whose temporary evolution supposes visualization.
  In this Letter we will map the solution of the equations of motion for a spin qudit  vector  on the three-dimensional oriented spherical
curve and show quantum interference of the precession  and nutation \cite{D3Viz}  to use these results for finding quantum speed limits (QSLs). \\
\indent We find QSLs  in the magnetic resonance %\cite{QSL}
as fundamental bounds
at minimum time which  is necessary for  a quantum system to
evolve into a different state. The first studies are the Mandelstam-Tamm, Margolus-Levitin-Toffoli \cite{MandelstamTamm,AnandanAharonov, Vaidman,MargolusLevitin,LevitinToffoli}
bounds, which have led to numerous extensions \cite{GeneralizedGSLs}. QSLs have many applications,
for instance, in a quantum cryptography \cite{MolotkovNazin}, quantum control \cite{OptimalControl,TECHNOLOGY}, quantum computation \cite{NatureMarkov},
communication \cite{Vedral}, quantum thermodynamics \cite{Deffner2010Lutz} and quantum metrology \cite{Plenio}. \\
 \section{Master equation}\label{II}
 We will restrict ourselves to a closed quantum system,
so that our initial state $\rho_0$ undergoes a unitary evolution and use the explicit model (1) at $h_1=h_2=h$.
 The  Liouville - Neumann equation for the density matrix $ \rho $,
describing the qudit dynamics looks like
%\end{document}
\begin{equation}\label{eq:2}
\partial_t\rho =-\imath [\hat {H}, \rho] ,
\rho (t=0) = \rho_0.
\end{equation}
The Hamiltonian of a magnetic qudit (spin S particle)
which is in the external  consistent  magnetic field
$ \overrightarrow {h (t)} $  is equal to
$\hat{H}=\mu C_i h_i(t)$,
where $h_i (t) $ are the Cartesian components of the  magnetic field in the frequency units,\,\,% (throughout this paper we take  $ \hbar=1$),
 $\mu$  is the qudit magnetic  moment; $C_1, \, C_2, \, C_3$ are  the matrix representation  of components  of the  qudit spin \cite{JMPIvanchenko}.
  We use  units chosen so,  that the magnetic moment $\mu$ equals 1 and $ \hbar=1$.
\\
The solution of the equation (\ref{eq:2}) looks like
\begin{equation}\label{eq:3}
    \rho =U.\rho_0.U^+,
\end{equation}
where $U=\alpha^{-1} e^{-\imath  t h_0}, \,\hat{h}=\alpha.\hat {H}.\alpha^{-1}-\imath \alpha.\partial_t \alpha^{-1}$.\, $\alpha=\mathrm {diag \,}(f(S),f(S-1),...,f(-S))$ is the $2S+1\times2S+1$ diagonal matrix, in which $f(S)=\mathrm {cn \,}(S\omega t|k)+\imath\, \mathrm {sn \,}(S\omega t|k)$  \cite{JMPIvanchenko,IJQI}, the Hamiltonian $\hat{h}$ is  independent of time $\forall$ $ S$ at $k=0$ and at $k\neq0$ for $S=1/2,1$.\\
It is convenient to rewrite Eq.~(\ref{eq:3})
 presenting the density matrix $%
\rho $ in the decomposition with the full set \cite{JMPIvanchenko} of
orthogonal Hermitian matrices $C_{\nu }$:
\begin{equation} \label {eq:4}
  \rho = \propto R _ {\nu} C_{\nu},
  ~  \rho ^ + =\rho, ~ \mathrm {Tr \,} \rho=1, ~R _ {0} =1,
\end{equation}
%\end{document}
in which, here and further, we imply the summation on  repetitive Greek
coefficients from zero to $(2S+1)^2-1$ and on Latin ones from one to three. The coherence  Bloch vector  $R _ {\nu}$  is widely used in the theory of the
magnetic resonance and characterizes the qudit behavior.
 In the case of the consistent field, as the expansion
 of the Rabi model,
  the solution (with the initial matrix  elements $\rho_{1,1} (t=0)=1$ and other ones are equal to zero)   at resonance $\omega=H$ for the spin components of  qubit or qutrit is the following:
\begin{equation}\label{eq:5}
\vec{R} = r_B( \mathrm {sn} (\omega t|k) \sin h t,
- \, \mathrm {cn} (\omega t|k) \sin h t, \cos h t),
\end{equation}
%\begin{eqnarray} \label {eq:5}
%R_1 &=& r_B\, \mathrm {sn} (\omega t|k) \sin h t, R_2 =
%- r_B\, \mathrm {cn} (\omega t|k) \sin h t,  R_3 = r_B\, \cos h t,
%\end{eqnarray}
where $r_B=\sqrt{3S/(S+1)}$ is the Bloch sphere radius. For  higher spins, the direct calculation looks the same (\ref{eq:5}) only if $k=0$.
\\
It will be useful to map
the solution for the qudit spin  vector on the geometrical
model \cite {FeynmanVernonHellwarth}. We parameterize a unit polarization vector by spherical angles $ (p_1, p_2, p_3) = (\sin\theta
\cos\varphi, \sin\theta \sin\varphi, \cos\theta)$.
Thus, the parameter $\varphi $ (0$ \leq\varphi\leq 2\pi $) becomes a precession angle
of the end of the vector $ \vec {p} $ on a sphere, referred to as an
apex, and the angle $ \theta $ (0$ \leq\theta\leq \pi $)
characterizes the nutation. And $ \theta=0$ corresponds to the north pole on the  sphere.
 In the resonance case  the nutation angular velocity has
the piecewise-constant time dependence
and the precession one is constant \cite{D3Viz}:
\begin{equation} \label {eq:6}
\theta \,' (\delta=0) = h \mathrm {sgn \,}( \sin h t), \,
\varphi \,' (\delta=0) = \omega
   \end{equation}
with the period $T=2 \pi/h $. \\
\indent In the consistent Rabi field  (\ref{eq:1}) at $ k\neq 0$, at the resonance, the angular velocities depend on  time and are equal to \cite{D3Viz}
%\end{document}%
\begin{eqnarray} \label {eq:7}
\theta _ {cf} \,' (\delta=0) &=& h \mathrm {sgn \,}(\sin h t),
\varphi _ {cf}\,' (\delta=0) = \omega \mathrm {dn \,} (\omega t|k).
\end{eqnarray}
\section{Minimum time for the evolution to the orthogonal
quantum state}\label{III}
The  variance of the qudit energy is read
 \begin{equation}\label{eq:8}
(\Delta
   E(t))^2=\mathrm {Tr \,} \rho \hat {H}^2-(\mathrm {Tr \,} \rho \hat {H})^2,
\end{equation}
where  $ \Delta E(t)$ is a standard deviation.\\
      The determination of the hodograph length  $s$ for pure states  \cite{AnandanAharonov} is
   \begin{equation}\label{eq:9}
    s=2 \int_0^t \Delta E(t') dt'
   \end{equation}
    and should be co-ordinated  with  the length on the sphere
    \begin{equation}\label{eq:10}
        l =\int_0^t \mathrm {v}d\tau,
    \end{equation}
    where the scalar of the vector velocity $\mathrm {v} =r_B\sqrt {p_1'^2+p_2'^2+p_3'^2}$ \cite{Aminov}, the prime is used to denote the derivative with respect to time, $ '\equiv\partial_t $.
    %\end{document}
    Having compared $s $ and $l$ from (\ref{eq:9}) and (\ref{eq:10}), we receive the formula
       \begin{equation}\label{eq:11}
    r_B\sqrt{p_1'^2+p_2'^2+p_3'^2}= \sqrt{4 p\,(\mathrm {Tr \,}  \rho \hat {H}^2-(\mathrm {Tr \,} \rho \hat {H})^2)},
   \end{equation}
     from which it follows,
that the velocity   of the apex is proportional  to the standard deviation  of the qudit energy. In this formula $p=1$ only for  spin $S =1/2 $  as the Fubini-Study metric is determined up to  a numerical factor \cite{StudyVinogradov,DodonovManko,GeneralizedGSLs}. The introduction of the parameter $p$ is a speed normalization.
With the increase of  spin $S $ the Bloch sphere radius grows, but it is useful to note that $ \lim _ {S \to \infty} \, r_B=\sqrt{3}$, and the parameter $p=\frac{3}{2(S+1)}$ decreases.\\
\indent The circulation on the closed path  at $k=0$
is proportional to the qudit energy variance, as it appears from the formulas  (\ref{eq:5}),\,(\ref{eq:11})
 \begin{equation} \label {eq:12}
 C=\oint_{\Gamma} \vec {R} \, '\cdot d\vec {R} =
 4 p\int_0^{T}(\Delta   E(t))^2 dt=
   \frac{3 \pi  S }{S+1}(2 h+\frac{H^2}{h}),
 \end{equation}
 where $T$
 is the time of passage of the closed contour $\Gamma$.
  At the fixed operating parameters $h$ and $H$ with  the growth of  spin $S$ the circulation grows.\\
\indent In the resonant case $\forall \omega=H$ the distance from the  north pole to the south pole on the Bloch sphere   during $ \tau =\pi/h$  equals
\begin{equation}\label{eq:13}
  s= r_B \int_0^{\tau} \sqrt{h^2+H^2 \sin ^2 h t\,
   \mathrm {dn \,}^2 ( H t|k)} dt \geq \pi r_B.
\end{equation}
As  $h\rightarrow\infty $, hence
\begin {equation} \label {eq:14}
  \lim\limits  _ {h \to \infty}  \int_0 ^ {\tau} \sqrt {h^2 +H^2 \sin ^2 h t \, \mathrm {dn \,} ^2 (H t|k)}\, dt =\pi,
\end {equation}
the bottom bound is obtained  in Eq. (\ref{eq:14}).
In other words, in a wide   interval  of the consistent field forms ($\forall k\in[0,1] $ and for all resonance cases  $\omega=H$) for a qubit and a qutrit \textit {the universal value} $ \pi $ is  established. \\
  At $k=0$ and for the higher spins $S\geq 3/2$  the same result  follows as
\begin{equation}\label{eq:15}
\lim\limits _ {h\to \infty}  \int_0 ^ {\tau} \sqrt {h^2 +H^2 \sin ^2 h t }\, dt =\lim_{h\to \infty } \, \mathrm {E}(\pi,
   -H ^2/h^2)=\pi,
\end{equation}
%\end{document}
where  $\mathrm {E}(\phi|m) = \int_0 ^\phi (1-m\sin^2\nu) ^ {1/2} d\nu $
   is an elliptic integral of the second kind. Thus, the  minimum distance $\pi r_B$   is \textit {the universal value} because
   it is independent of field parameters in Eqs. (14,15).
   This  distance is nonzero even though $\tau \to 0$ as $h\to \infty $.\\
\indent The evolution     of the initial  qudit spin vector
 with the doubly stochastic  matrix
     and the diagonal time-independent  Hamiltonian
     $ \eta\,C_3$ \cite {AnandanAharonov}
     geometrically presented on the Bloch sphere is a precession in the equatorial plane:
      $ (R_1, R_2, R_3) = r_b(
\cos\eta t, \sin\eta t, 0) $, $ \varphi \, ' (t) = \eta $, where $ \eta $ is
   a positive constant. This  geodesic line has the length  $\pi r_b$ during half - time $\pi/\eta$. In this case from Eq. (\ref{eq:11}) follows  the formula %$ \,2/\sqrt{3}   \sqrt{p\, S(S+1)} =  \,r_B$.
    $    p =\frac{3r_b^2}{4   \, S(S+1)}$.
   The straightforward calculation shows that as well as in  case of the
   variable magnetic field, the   Bloch sphere radius $r_b$  is finite
   for  qudit spins:
    $(S,r_b,p)=(1/2,1,1),(1,1.15,1/2),
   (3/2,1.22,0.3),(2,1.26,0.2),
   %(\frac{5}{2},1.28,0.14)$,
   %$(3,1.3,0.10),(\frac{7}{2},1.31,0.08)
   ...,\\(4,1.32,0.06)
   $,...,$(10,1.86,0.02),...\, .$
   Therefore at the infinitesimal time the apex
    passes the infinitesimal distance in this model. The distance for large spins, incompatible with   the distance on the sphere Eq. (10),  was used in (\cite{AnandanAharonov}, Eq. (15)).\\
  \indent We have from (\ref{eq:11})
  \begin{equation}\label{eq:16}
  \int_0^{\frac{\pi}{h}}  \Delta
   E(t) dt \geq \frac{\pi r_B}{\sqrt{4 p}}= \sqrt{\frac{S}{2}}\pi.
\end{equation}
Hence the Mandelstam-Tamm relation \cite {MandelstamTamm, deffner2013energy} between the averaged on time $\tau$  standard deviation energy
$ \Delta E_{\tau }=\tau^{-1}\int_0 ^ {\tau} \Delta   E (t) dt $ during  $ \tau=\pi/h$ becomes
     \begin{equation}\label{eq:17}
\Delta E_\tau \,\tau \geq  \sqrt{\frac{S}{2}}\pi.
   \end{equation}
   The bottom bound $\tau_{QSL}$ is determined by the formula
       \begin{equation}\label{eq:18}
   \tau \geq
\frac{\pi ^2 \sqrt{S}}{ \sqrt{2} h \mathrm {E}\left(-H ^2/h^2\right)}=\tau_{QSL}.
      \end{equation}\\
   At $H\ll h$ the bottom bound goes to zero for $ \forall S$; at $H\gg h$ the non-nil bottom bound  $\lim\limits _ {h\to 0} \tau_{QSL}=\frac{\pi ^2 \sqrt{S}}{\sqrt{2} H} $ is attained.\\
      \indent The distance from the north pole on the Bloch sphere  to the nearest orthogonal state
            is more or equal
      $ \pi r_B/(2S)$
   in a qudit during $\tau_1=\pi/(2Sh)$, because  at a resonance  the nutation speed  has the piecewise-constant time dependence (\ref{eq:6}), (\ref{eq:7}).
     %
      %(see Fig. 5,\,6 \cite{D3Viz})
       At $h\rightarrow\infty$ the  value  is $ \pi/(2S) $ as it follows from  (\ref{eq:15}). The bottom bound in this case goes down in comparison with a qubit.\\
   \indent The  relation  between  the time-averaged  standard deviation energy  $ \Delta E_{\tau_1} =\tau_1^{-1} \int_0 ^ {\tau_1} \Delta
   E (t) dt $ during  $ \tau_1 $ is
      \begin{equation}\label{eq:19}
\Delta E_{\tau_1} \,\tau_1 \geq %\frac{\pi\sqrt{2S}}{4S}=
\frac{\pi}{2\sqrt{2S}}.
   \end{equation}
   The  equation (\ref{eq:19}) directly leads to
     \begin{equation}\label{eq:20}
   \tau_1 \geq\frac{\pi ^2}{2  h (2 S)^{3/2} \mathrm {E}\left(\pi/
   (2S),-H ^2/h^2\right)}={\tau_1}_{QSL},
   %\tau_1 \geq \frac{\pi}{2\sqrt{2S}  \Delta E_{\tau_1} },
   \end{equation}
   from which it is obvious that the bottom bound goes down at the use of multilevel spin systems. At $H\ll h$ the bottom bound goes to zero for $ \forall$ $ S$; at $H\gg h$  the  non-nil bottom bound  $\lim\limits _ {h\to 0} {\tau_1}_{QSL}=\frac{\pi ^2}{H (2S)^{3/2}( (-1)^{r[\frac{1}{2 S}]}(1-|\cos(\frac{\pi}{2S})|)+2r[\frac{1}{2 S}])} $ is attained. The function r[x]
gives the integer closest to x. The ratio $(\lim\limits _ {h\to 0} \tau_{QSL}/{\tau_1}_{QSL},S)$ is $(1,1/2),
(2,1),(9/4,3/2),(2.343,2),...,(2.467,S\gg1)$. Hence, there is a possibility  to use  multilevel systems for a  transition time reduction  between orthogonal states.\\
\indent
The  quantum speed limit
time is defined  with respect to a given class
of  Hamiltonians and  the
minimal time depends on  the driving parameters.
Our results  coincide qualitatively with the results \cite {HegerfeldtPRA}  for the whole class of spin Hamiltonians.
 We have shown that % for time-dependent Hamiltonians,
the QSL determination  requires the knowledge
of the  system state  during the evolution.
\subsection{Dynamics of  spin standard deviations}
\begin{figure}%[h]
\includegraphics [width = 2.2 in] {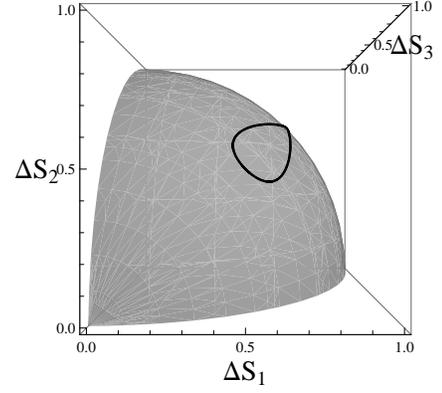}
\caption {\label {Fig_1_RSumvar0} The top view of the parametrical plot ($\Delta S_1(t),\Delta S_2(t),\Delta S_3(t)$) of the closed space curve  (\ref{eq:22})
%the positive octant of a 3 dimensional  sphere of unit radius.
 on the positive 1/8  part of the sphere for a spin 1 with the parameters of the circularly polarised field $ \omega=H=1$,
\, $ h =2$.}
\end{figure}
 \begin{figure}[h]
\includegraphics [width = 2.2 in] {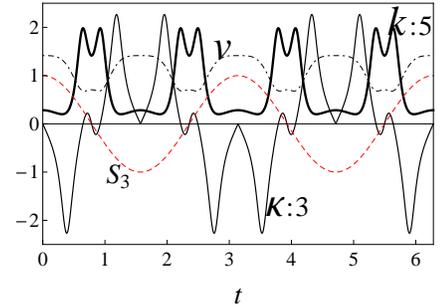}
\caption {\label {Fig_2_RViz} Dynamics of  $S_3$,  curvature $k$, torsion of  curve $\kappa$,
velocity modulus of an apex $V$  for a spin 1 with the parameters as in Fig. 1.}
\end{figure}
The real and non-negative eigenvalues
of the $ 3 \times 3 $ covariance matrix
\begin{equation}\label{eq:21}
\mathrm {Cov}(S_i,S_k)=\frac{1}{2}
   \mathrm {Tr \,}\left[\rho\,\left(C_iC_k+C_kC_i\right)\right]-\mathrm {Tr \,}\left[\rho\,C_i\right]
   \mathrm {Tr \,}\left[\rho\,C_k\right]
\end{equation}
 equal $\lambda_1=\lambda_2=\frac{S}{2},\lambda_3=0$ and the invariants  of the tensor  are $\lambda_1+\lambda_2, \lambda_1\lambda_2, \lambda_3$. \\
\indent Hence,
the  sum of the variances \cite {Pati,Fei} of the spin components $S_i $ in the  magnetic resonance in the circularly polarized magnetic field is conserved for any $t$ during the evolution:
\begin{equation}\label{eq:22}
(\Delta S_1)^2+(\Delta S_2)^2+(\Delta S_3)^2=S,
\end{equation}
where $
(\Delta S_1,\Delta S_2,\Delta S_3)=1/2\sqrt{S/2}(
\sqrt{3+\cos 2 h
   t+2 \sin ^2 h t \cos 2  \omega t},\\
\sqrt{3+\cos 2 h
   t-2 \sin ^2 h t \cos 2  \omega t},
   2\left| \sin h t\right|).$
In the  consistent field the relation (\ref{eq:22}) is fulfilled only for a qubit  and a qutrit.\\
  Due to $\Delta S_1(t)\geq 0,\Delta S_2(t)\geq 0,\Delta S_3(t)\geq 0$ and quantum restriction (22), the curve lies on 1/8 part of the sphere. This  curve is the closed one  with period $T_c=2\pi m/h=\pi l/\omega$, where $m, l$ are integers. The oriented spherical curve  is characterized by the curvature $k$, torsion of the curve $\kappa$,  apex speed  $V$, and length of the path $s$ \cite{Aminov,D3Viz}. In Fig. 2 it is shown that the speed  $V$ is minimum, when $S_3$ and  $\kappa$ change a sign and $k$ has a local minimum. $S_3$ is extreme, when $V$ is maximal, $k$ is minimum, $\kappa$  has a local maximum/local minimum. The quantitative characteristic of the curve at the coordinates $\Delta S_1,\Delta S_2,\Delta S_3$ in Fig. 1  represents in details the resonant evolution  in Fig.\,2.\\
\indent It is possible to apply a known bilateral inequality   (between the  harmonic, geometrical and arithmetic averages)
to the   estimation of the product of two $\Delta S_i\Delta S_k ,(i\neq k =1,2,3)$ and three standard deviations $\Delta S_1\Delta S_2\Delta S_3$.
%
%\begin{equation}\label{eq:23}
%    \frac{2}{\frac{1}{(\Delta S_i)^2}+\frac{1}{(\Delta S_j)^2}}\leq\Delta S_i\Delta S_j\leq\frac{(\Delta S_i)^2+(\Delta S_j)^2}{2},\,i\neq j =1,2,3,
%\end{equation}
%\begin{equation}\label{eq:24}
%(\frac{3}{\frac{1}{(\Delta S_1)^2}+\frac{1}{(\Delta S_2)^2}+\frac{1}{(\Delta S_3)^2}})^{\frac{3}{2}}\leq\Delta S_1\Delta S_2\Delta %S_3\leq(\frac{S}{3})^{\frac{3}{2}}.
%\end{equation}
\indent The dynamics of the spin standard deviations and their products   is presented in Fig. 1,3.
 \begin{figure}%[h]
\includegraphics [width = 2.2 in] {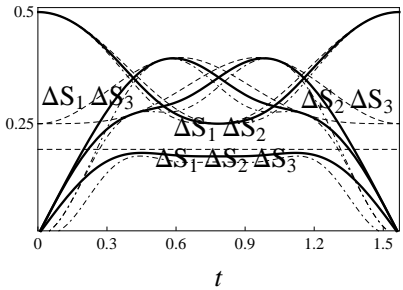}
\caption {\label {Fig_3_RProduct1} Dependence of the products of two and three standard deviations on time
%according to the formulae (\ref{eq:23}), (\ref{eq:24})
for a spin 1 with the parameters
%of the circularly polarised field
as in Fig. 1. %$ \omega=H=1$,\, $ h =2$.
Dashed and dot-dashed lines accordingly specify  the arithmetic average   and the  harmonic one; black thick lines show the analytic solution.}
\end{figure}
\begin{figure}
\includegraphics [width =2.2 in] {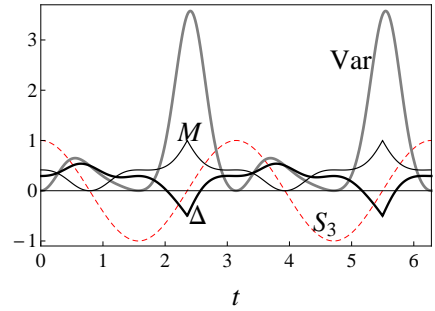}
\caption {\label {Fig_4_RVar} Dynamics  of $S_3,\,M(S_1:S_2),\,\Delta(S_1|S_2),\,\mathrm {Var \,}(S_1|S_2)$ for a spin 1 with the parameters
as in Fig. 1.}
\end{figure}
In Fig. 3 the plots of the  harmonious, geometrical and arithmetic averages  are presented. As it can be seen in Fig. 3 for both $\Delta S_1\Delta S_2$ and
$\Delta S_1\Delta S_3, \Delta S_2\Delta S_3$  the inequalities  practically transform locally into the equalities  when the equation (\ref{eq:22}) represents the closed  curve on the sphere.\\
\indent In terms of the standard spin deviations and new notions  \cite{MutualUncertainty}:
it is possible to describe the dynamics of the  magnetic resonance with help of the mutual uncertainty ($M$) $ M (S_i:S_k) = \Delta S_i +\Delta S_k-\Delta (S_i+S_k) $,
the conditional uncertainty ($ \Delta $) $ \Delta (S_i|S_k) = \Delta (S_i+S_k)-\Delta
 S_k $ and the conditional variance (Var) $ \mathrm {Var \,} (S_i|S_k) = \mathrm {Var
\,} (S_i+S_k)-\mathrm {Var \,} S_k $.\\
From Fig. 2,4,5 it is seen  that  M, $ \Delta $, Var are extreme when $S_3$ changes sign. For the spin $S \geq 1$
we obtain $M\geq 0$, $\mathrm {Var \, }\geq 0$  and $ \Delta $ is sign-changing. The conditional variance equals zero in the exact resonance ($S_3=-1$). These new characteristics specify the description of the  magnetic resonance.
\begin{figure}
\includegraphics [width =2.2 in] {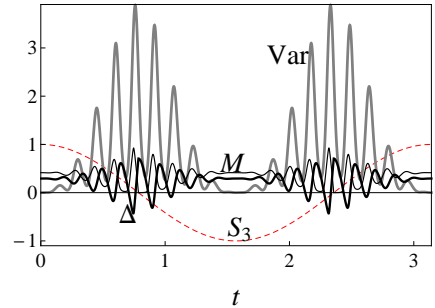}
\caption {\label {Fig_5_RVar1} The same as in Fig. 4 with the parameters $\omega=H=20$,\, $ h =2$.}
\end{figure}
The description of the magnetic resonance both  in  geometrical terms of Figs. 1,2 and in terms of  conditional uncertainty and variance Figs. 4,5 mutually complement each other.
\section{Conclusion} \label{IV}
At resonance  in the circularly polarization field
the Bloch sphere radius  is limited by value $ \sqrt {3}$ when $S\rightarrow\infty$.
 The  Fubini-Study measure  in the finite-dimensional spin space is specified. The universal  length value    is found both for $S=1/2,1$ in  the coordinated magnetic field  and for any $S$  in the circularly polarization field.
 For any  spin the bottom bounds QSLs are found.
 With the spin growth the transition time  between levels decreases.
  The minimal time goes to zero at $H\ll h$.
The dynamics of  the spin standard deviations has been presented.
  An experimental confirmation of our  results
 can be implemented  at $H\gg h$ using a NMR setup \cite{JMR,ExperimentalDemonstration}.\\
%\indent This research did not receive any specific grant from funding agencies in the public, commercial, or
%not-for-profit sectors .
\section*{Acknowledgment}
 The author is thankful to  Yu. L. Bolotin for useful discussion.
%\section*{References}
% \bibliographystyle {plain}
% \bibliographystyle {unsrt}
%\bibliographystyle {apsrev}
% \bibliographystyle {gost780u}
%%\bibliography{Sphera_a,sphera_b}
%\bibliography{Sphera_a}
%\bibliography{sphera_b}
%\end{document}
%
\end{document}